\documentclass[twocolumn]{aastex7}
 
\usepackage{comment}
\usepackage{enumitem}
\usepackage{booktabs}
\usepackage{tabularx}
\usepackage{amsfonts,amsmath,amssymb}

\newcommand{\mgii}{Mg{\sc~ii}}
\newcommand{\ciii}{C{\sc~iii}}
\newcommand{\heii}{He{\sc~ii}}
\newcommand{\siiv}{Si{\sc~iv}}
\newcommand{\ovi}{O{\sc~vi}}
\newcommand{\oiii}{O{\sc~iii}}
\newcommand{\oii}{O{\sc~ii}}
\newcommand{\nv}{N{\sc~v}}
\newcommand{\niii}{N{\sc~iii}}
\def\civ{C\,{\sc~iv}}

\newcommand{\ergs}{${\rm erg \ cm^{-2} \ s^{-1}}$ }

\begin{document}

\title{Revealing the Physical Driver of the Baldwin Effect: Gas Density in the Broad‑Line Region}

\correspondingauthor{Zhi-fu Chen}
\email{zhichenfu@126.com}

\author[orcid=0009-0001-3122-3237]{Wen-Qiang Liang}
\affiliation{School of Physics and Electronic Information, Guangxi Minzu University, Nanning 530006, China}
\affiliation{Guangxi Key Laboratory for Relativistic Astrophysics, School of Physical Science and Technology, Guangxi University, Nanning 530004, China}
\email{liangwenqiang@st.gxu.edu.cn}  

\author[orcid=0000-0003-0639-1148]{Zhi-Fu Chen} 
\affiliation{School of Physics and Electronic Information, Guangxi Minzu University, Nanning 530006, China}
\email{zhichenfu@126.com}

\author[orcid=0000-0003-2467-3608]{Rui-Jing Lu}
\affiliation{Guangxi Key Laboratory for Relativistic Astrophysics, School of Physical Science and Technology, Guangxi University, Nanning 530004, China}
\email{luruijing@gxu.edu.cn}

\author{Cheng-Feng Peng}
\affiliation{Guangxi Key Laboratory for Relativistic Astrophysics, School of Physical Science and Technology, Guangxi University, Nanning 530004, China}
\email{2507401013@st.gxu.edu.cn}

\author{Shao-Hua Zhang}
\affiliation{Shanghai Key Lab for Astrophysics, Shanghai Normal University, Shanghai 200234, China}
\email{zhangshaohua@shnu.edu.cn}

\author{Zhe-Geng Chen}
\affiliation{School of Physics and Electronic Information, Guangxi Minzu University, Nanning 530006, China}
\affiliation{Guangxi Key Laboratory for Relativistic Astrophysics, School of Physical Science and Technology, Guangxi University, Nanning 530004, China}
\email{chenzhegeng@st.gxu.edu.cn}

\author{Zhi-Qing Chen}
\affiliation{School of Physics and Electronic Information, Guangxi Minzu University, Nanning 530006, China}
\email{1226332980@qq.com}

\author{Yi-Qi Chen}
\affiliation{School of Physics and Electronic Information, Guangxi Minzu University, Nanning 530006, China}
\email{13008314050@163.com}

\author{Dong-Li Jiang}
\affiliation{School of Physics and Electronic Information, Guangxi Minzu University, Nanning 530006, China}
\email{}

\author{Xi-Heng Shi}
\affiliation{MNR Key Laboratory for Polar Science, Polar Research Institute of China, Shanghai 200136, China}
\email{shixiheng@pric.org.cn}

\author{Zhi-Wen Wang}
\affiliation{School of Physics and Electronic Information, Guangxi Minzu University, Nanning 530006, China}
\email{wzhiwen2004@126.com}

\begin{abstract}
The Baldwin effect --- the inverse correlation between the equivalent width of emission lines and the continuum luminosity in active galactic nuclei (AGNs) --- has been known for nearly five decades, yet its physical origin remains poorly understood. Using a sample of 41,159 radio quasars constructed from the Sloan Digital Sky Survey and the Low-Frequency Array Two-metre Sky Survey, we investigate the origin and underlying physics of the Baldwin effect of \mgii\ broad emission lines in both radio-quiet (RQ) and radio-loud (RL) quasars. We find that the slope $\beta$ of the Baldwin effect is positively correlated with the Eddington ratio $\lambda_{\rm Edd}$ in both populations, and RL quasars exhibit steeper $\beta$ than their RQ quasars at fixed $\lambda_{\rm Edd}$. Photoionization simulations reveal that the $\beta$ is primarily governed by the gas density in the broad-line region (BLR): lower gas densities yield steeper slopes. This density‑driven mechanism naturally connects the Baldwin effect to the broader AGN evolutionary context. Specifically, higher $\lambda_{\rm Edd}$ drive stronger accretion disk winds, leading to denser BLRs and shallower $\beta$. Our findings indicate that BLR gas density serves as the primary physical driver underlying the  ``global" Baldwin effect, offering a physically grounded framework for interpreting AGN accretion states and their coupled evolution with host galaxies.
\end{abstract}

\keywords{galaxies:active --- galaxies:jets --- quasars:emission lines}

\section{INTRODUCTION} \label{sec:intro}
The Baldwin effect was first reported as an inverse correlation between the equivalent width (EW) of broad \civ\ emission lines and the continuum luminosity at 1350 \AA\ \citep{Baldwin_1977}. It has since been observed in other broad lines (e.g., \ovi\ $\lambda$1034, Ly$\alpha$, \nv\ $\lambda$1240, \siiv\ $\lambda$1400, \heii\ $\lambda$1640, \ciii] $\lambda$1909, \mgii\ $\lambda$2800, H$\beta$ $\lambda$4861) (e.g., \citealt{Wampler_1984, Baldwin_1989, Boroson_1992, Netzer_1992, Francis_1995, Baskin_2004}), in narrow lines (e.g., \oiii $\lambda$5007, \oii $\lambda$3727, [\nv] $\lambda$3426, [\niii] $\lambda$3869) (e.g., \citealt{Laor_1995, Croom_2002, Netzer_2006}), and even in the $\rm Fe~K\alpha$ X-ray line (e.g., \citealt{Iwasawa_1993, Green_1996}). The effect is present both in ensemble samples (``global'' Baldwin effect) and in individual, variable AGNs (``intrinsic'' Baldwin effect) (e.g., \citealt{Pogge_1992, Goad_2004, Rakic_2017}). Meanwhile, the slopes of the Baldwin effect have been found to steepen with increasing ionization potential of the emission lines (e.g.,\citealt{Espey_1999, Dietrich_2002, Zhang_2013}). 

Despite extensive observational efforts, the physical mechanism underlying the Baldwin effect remains elusive. Under standard photoionization theory \citep{Osterbrock_2006}, the line luminosity scales linearly with continuum luminosity, implying that the slope ($B$) of the line–continuum luminosity relation and the Baldwin slope ($\beta$) satisfy $B-\beta=1$. Some studies suggest that the Eddington ratio (the ratio between the bolometric and Eddington luminosities: $\lambda_{\rm Edd} = L_{\rm bol}/L_{\rm Edd}$) is a key driver, with a tight correlation found for certain lines (e.g., \citealt{Bachev_2004, Baskin_2004, DongXB_2009, Bian_2012}), while others emphasize the role of black hole mass (e.g., \citealt{Warner_2003, Warner_2004, Xu_2008}). In particular, the \mgii\ EW appears to be independent of luminosity and primarily governed by $\lambda_{\rm Edd}$ \citep{DongXB_2009}, leading to the proposal that the classical \mgii\ Baldwin effect is a secondary effect arising from the EW–$\lambda_{\rm Edd}$ relation combined with flux-limited sample selection. However, this interpretation has been challenged by studies of narrow lines (e.g., \citealt{Zhang_2013}). Radio loudness has also been considered as a potential influencing factor, with some works reporting a steeper Baldwin slope in RL AGNs compared to RQ ones (e.g., \citealt{Osmer_1994, Victor_2016, Xiao_2022}), while others find no significant difference (e.g., \citealt{Kinney_1990}). A consensus has yet to emerge.

In this study, we aim to disentangle the respective roles of Eddington ratio and radio-loudness in governing the ``global" Baldwin effect of \mgii\ broad emission lines, and to explore whether a unified physical framework can explain their interplay. The structure of this paper is organized as follows. In Section \ref{sec:data}, we describe the sample selection of quasars. The results and analysis are presented in Section \ref{sec:results}. Finally, we discuss the implications of our findings and make a brief summary in Section \ref{sec:discussion}. Throughout this study, we adopt a flat $\Lambda$CDM cosmology with $\Omega_m = 0.3$, $\Omega_{\Lambda} = 0.7$, and $H_0 = 70\ \rm{km}\ \rm{s}^{-1}\ \rm{Mpc}^{-1}$.

\section{Quasar sample} \label{sec:data} 
The Sloan Digital Sky Survey (SDSS) employed a dedicated 2.5-meter wide-field telescope \citep{Gunn_2006} to acquire optical quasar spectra. SDSS-I/II \citep{Abazajian_2009} obtained spectra spanning 3800--9200 \AA\ at a spectral resolution of $R \approx 2000$, while SDSS-III/IV \citep{Dawson_2013, Smee_2013, Dawson_2016} extended coverage to 3600–10,500 \AA\ with $R \approx 1300 \sim 2500$. The Sixteenth Data Release Quasar Catalog (DR16Q) \citep{Lyke_2020} consolidates spectroscopic identifications from all four SDSS phases, comprising 750,414 confirmed quasars. Building upon DR16Q, Wu et al. \citep{Wu_2022} derived a comprehensive catalog of continuum and broad-emission-line properties—including rest-frame equivalent widths, luminosities, and line-profile parameters—thereby enabling robust statistical investigations of the Baldwin effect.

The LOw-Frequency ARray (LOFAR, \citealt{Shimwell_2017}) Two-metre Sky Survey (LoTSS, \citealt{van_Haarlem_2013}) provides high-fidelity, sub-arcsecond radio imaging of the northern sky, achieving an angular resolution of 6$''$ at its central observing frequency of 144 MHz (bandwidth: 120–168 MHz). Designed to probe galaxy evolution across cosmic time, the LoTSS detects synchrotron emission associated with AGN jets and star formation activity. The second data release (LoTSS-DR2, \citealt{Shimwell_2022}) contains calibrated radio measurements for 4,396,228 sources. Critically, the LoTSS DR2 exhibits substantial sky coverage overlap with both the SDSS imaging and spectroscopic footprints, rendering it especially well-suited for identifying radio counterparts to optically selected SDSS quasars.

We adopt the radio-optical crossmatch catalogue of \citealt{Hardcastle_2023}, which uses a likelihood-ratio method (e.g., \citealt{Sutherland_1992,Williams_2019}). From this catalogue, we select 53,621 SDSS DR16Q quasars with clean SDSS redshift (${\rm zwarning}=0$) and successful optical identification (${\rm ID\_flag}>0$) as our parent sample. 
To investigate the Baldwin effect in the \mgii\ broad emission line, we restricted our analysis to quasars within the redshift range  $z = 0.385 \sim 2.467$—ensuring \mgii\ falls within the SDSS spectroscopic window—and required median signal-to-noise ratios (S/N) $>2$ to ensure reliable spectral fitting (e.g., \citealt{Huang_2023}). We further applied rigorous quality cuts to exclude objects with poorly constrained measurements: (1) \mgii\ EW exceeding 3 $\sigma_{\rm EW}$; (2) rest-frame 3000 \AA\ continuum luminosity $L_{\rm 3000} > 10^{43}$ \ergs, ensuring sufficient luminosity for meaningful Baldwin-effect analysis; (3) logarithmic uncertainty in $L_{3000} < 0.2$ dex, guaranteeing precision in the luminosity measurement; and (4) all physical quantities were required to be positive-definite. After applying these criteria, our final working sample comprises 41,159 quasars.

\section{Results} \label{sec:results}
\subsection{Radio loudness and Eddington ratio}  \label{subsec:method}
The radio-loudness parameter is defined as $R = \log (L_{\rm{1.4\,GHz}}/L_{\rm{i}})$, where $L_{\rm{1.4\,GHz}}$ and $L_{\rm{i}}$ denote the monochromatic luminosity at 1.4 GHz and the integrated luminosity in the optical $i$-band, respectively. 
The 1.4 GHz luminosity is derived from the integrated 144 MHz flux density ($S_{\rm 144MHz}$) reported in the LoTSS DR2, assuming a canonical radio spectral index of $\alpha_{\rm rad} = -0.7$ (i.e., $f\propto \nu^{\alpha_{\rm rad}}$). Specifically, the monochromatic luminosity at 1.4GHz is computed as: 
\begin{equation}\label{L1.4Hz}
    L_{\rm 1.4GHz}=4\pi d^{2}_{L} S_{\rm 144MHz} (\frac{1400}{144})^{\alpha_{\rm rad}} (1+z)^{-(1+\alpha_{\rm rad})},
\end{equation}
where $d_{L}$ is the luminosity distance. 
For the $i$-band luminosity $L_{\rm{i}}$, we first compute the absolute $i$-band magnitude at redshift $z = 0$ ($M^{\rm{z=0}}_{\rm{i}}$) using the relation:  
\begin{equation}\label{magnitude correlation}
	M^{\rm{z=0}}_{\rm{i}} = M^{\rm{z=2}}_{\rm{i}} + 2.5(1 + \alpha_{\rm{opt}})\rm{log_{10}}(1 + z),
\end{equation}
where $M^{\rm{z=2}}_{\rm{i}}$ represents the $k$-corrected absolute $i$-band magnitude at z = 2 from the SDSS-IV DR16Q, and an optical spectral index of $\alpha_{\rm{opt}} = -0.5$ is adopted throughout this work. Subsequently, $L_{\rm{i}}$ is calculated via the standard luminosity–magnitude relation, with the Sun serving as the photometric reference.

\begin{figure*}
	\centering
	\includegraphics[scale=0.4]{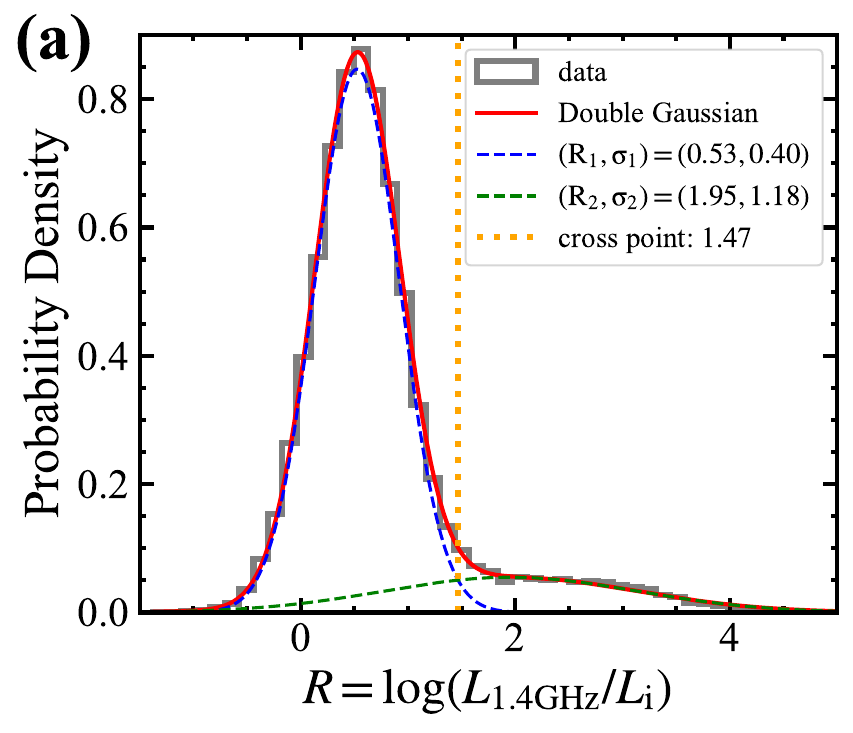}
	\includegraphics[scale=0.4]{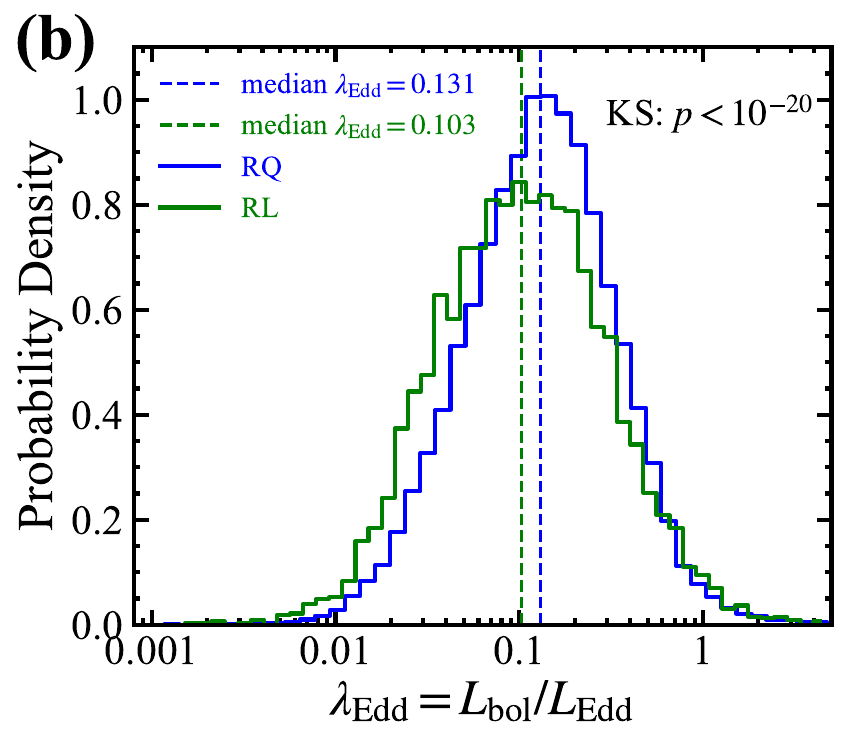}
	\includegraphics[scale=0.4]{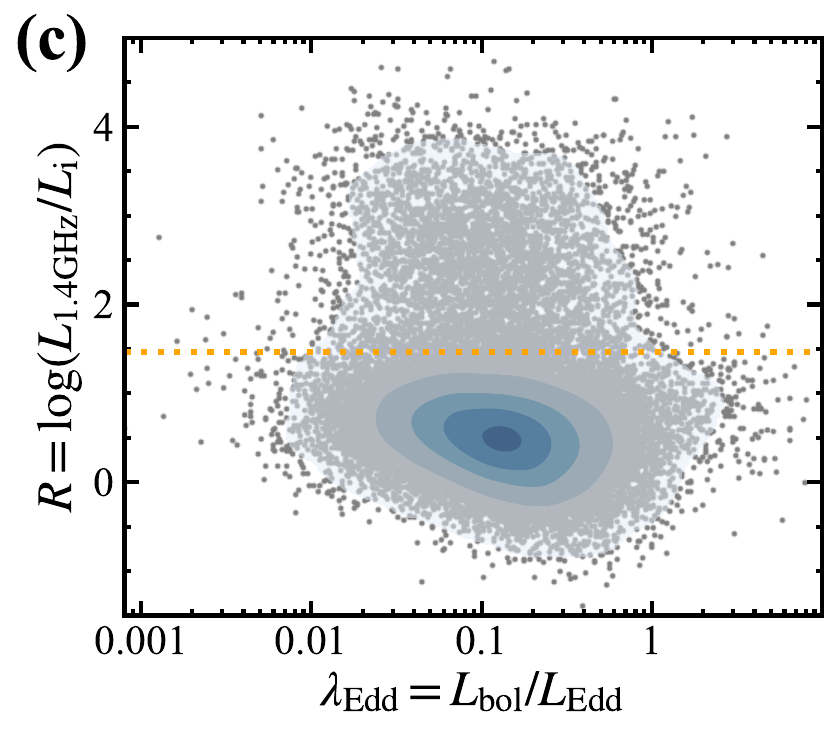}
	\caption{(a) The Radio-loudness distribution (the gray stepped line) of our sample is decomposed by fitting two Gaussian functions (the blue and green dotted lines), and the red solid line represents the sum of the two components. The intersection point ($R=1.47$) of the two components is indicated by the vertical dotted line. (b) Eddington ratio distributions for RQ (blue solid line) and RL (green solid line) quasars. RQ and RL quasars have median $\lambda_{\rm Edd}=0.131$ and $\lambda_{\rm Edd}=0.101$, respectively. The Kolmogorov–Smirnov test ($p < 10^{-20}$) confirms that the two distributions are statistically distinct. (c) Radio-loudness versus Eddington ratio. The orange dotted line indicates the boundary between the two types of quasars. The shadow represent the kernel density estimates illustrating the distribution trends of the data points. The Pearson correlation analysis yields weak correlations between $R$ and $\lambda_{Edd}$ for both RQ and RL quasars: correlation coefficients of $r = -0.20$ ($p < 10^{-22}$) and $r = -0.09$ ($p < 10^{-20}$) for RQ and RL quasars, respectively.}
	\label{fig:lambda and R}
\end{figure*}

As illustrated in Figure {\ref{fig:lambda and R}a}, the distribution of the resultant radio-loudness parameter $R$ for our quasar sample exhibits a  pronounced bimodal structure. A double-Gaussian function fit to this distribution yields an intersection point at $R = 1.47$, which serves as the empirical threshold for distinguishing RQ from RL quasars. Based on this criterion, the sample is classified into 36,578 RQ quasars ($R \leq 1.47$) and 4,581 RL quasars ($R > 1.47$). 
Their corresponding Eddington ratios\footnote{Black hole masses in Wu \& Shen (2022) are estimated via the virial method using H$\beta$/Mg\,II/C\,IV depending on redshift, with $L_{\rm bol}\approx5L_{3000}$ adopted for the bolometric correction. For our sample restricted to $z=0.385$--$2.467$, the Eddington ratios are thus Mg\,II-based.} (taken from \citealt{Wu_2022}) are presented in Figure \ref{fig:lambda and R}b, revealing median values of 0.131 for the RQ sample and 0.101 for the RL sample.
The extremely small $p$-value from the Kolmogorov–Smirnov test ($p < 10^{-20}$) confirms that the Eddington ratio distributions of the RQ and RL subsamples are statistically distinct. 

The $\lambda_{\rm Edd}$–$R$ plane is also presented in Figure \textcolor{blue}{\ref{fig:lambda and R}c}. Pearson correlation analysis yields a correlation coefficient of $r = -0.20$ ($p < 10^{-22}$) for RQ quasars, and $r = -0.09$ ($p < 10^{-20}$) for RL quasars. These results indicate that, although the correlations between the Eddington ratio and radio loudness are highly statistically significant ($p \ll 10^{-4}$) in both RQ and RL quasar subsamples, the strength of the correlations remains relatively weak ($|r| < 0.3$).

\subsection{Baldwin effect of \mgii\ emission lines}  \label{subsec:BEff}
\begin{figure*}
\centering
\includegraphics[scale=0.35]{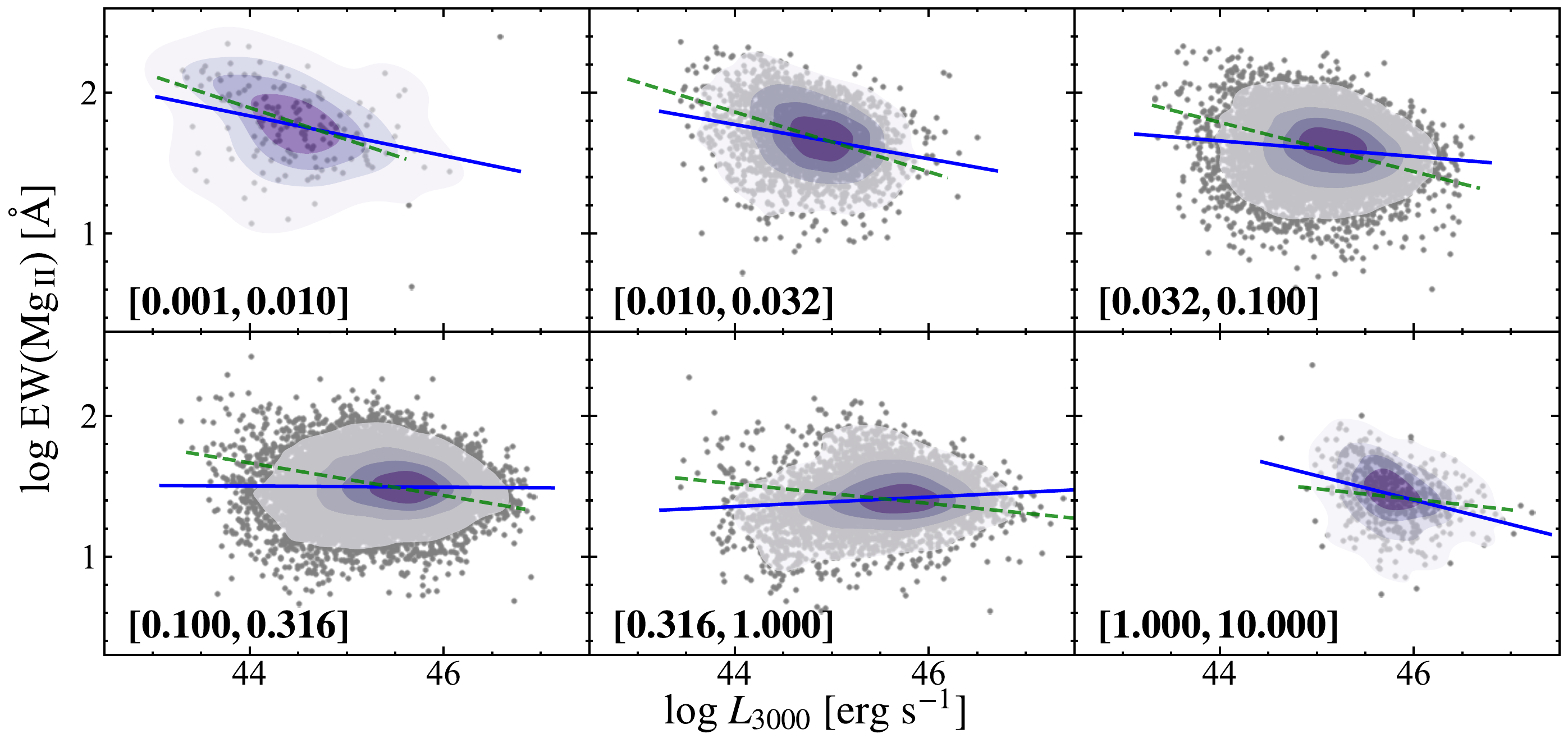}
\caption{Correlation between the \mgii\ EW and the rest-frame continuum luminosity at 3000  \AA\ ($L_{3000}$) for RQ quasars, binned by $\lambda_{\rm Edd}$, and the $\lambda_{\rm Edd}$ range for each bin is indicated in the lower-left corner of the corresponding panel. Blue solid lines show the best-fitting power-law relations, purple filled contours represent kernel density estimates, illustrating the underlying data distribution. For a comparison, the best-fit lines for the RL quasars are overplotted as green dashed lines.}
\label{fig:BE_RQ}
\end{figure*}

\begin{figure*}
	\centering
	\includegraphics[scale=0.35]{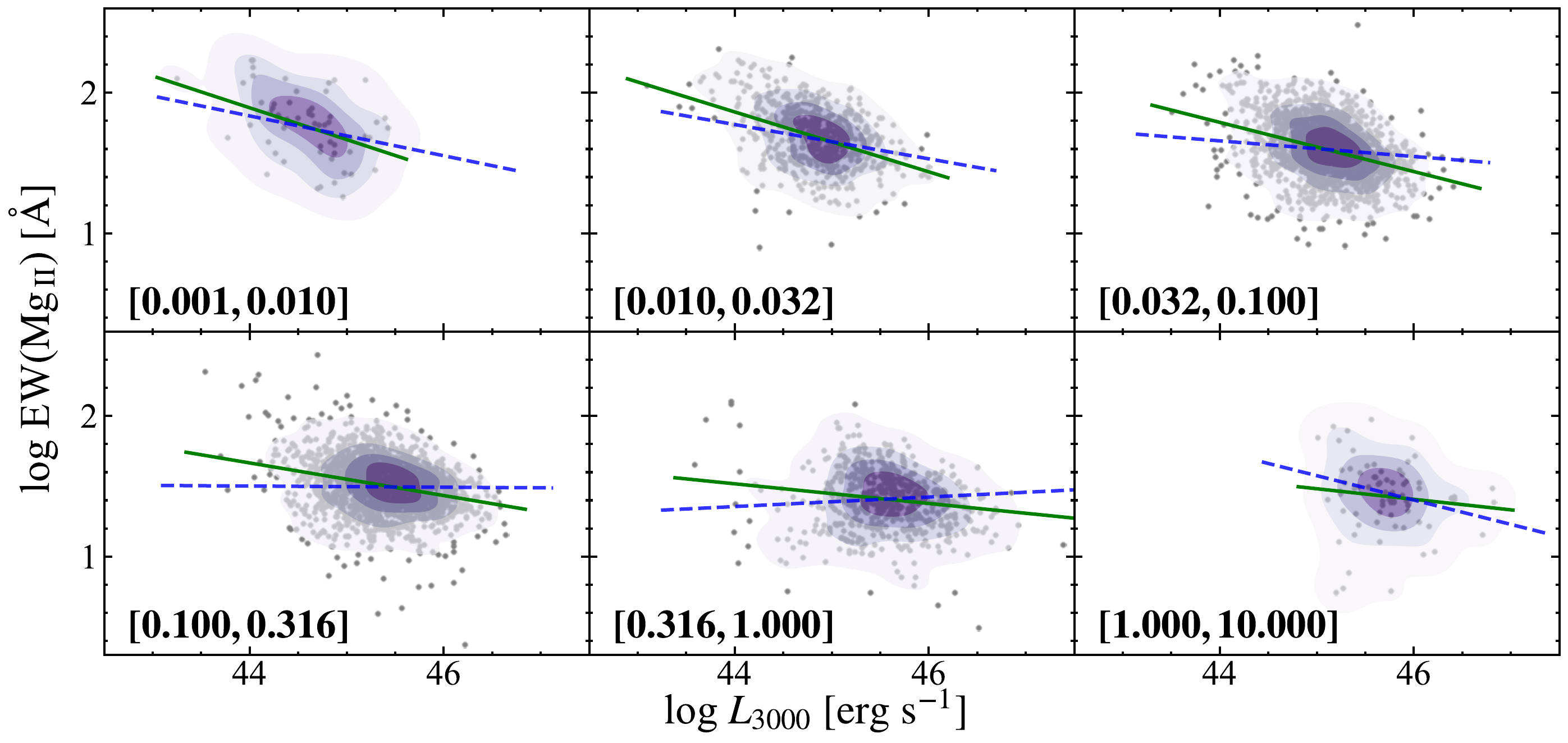}
	\caption{The same as Figure \ref{fig:BE_RQ} but for RL quasars. The green solid lines denote the best-fit Baldwin relations from the RL quasars, while the blue dashed line are also overplotted for comparison. }
	\label{fig:BE_RL}
\end{figure*}

To investigate whether the Baldwin effect manifests differently between RL and RQ quasars as a function of $\lambda_{\rm Edd}$, we separately binned each quasar sample into several bins, in ascending order of Eddington ratio. For each $\lambda_{\rm Edd}$ bin, we perform a linear regression of the form: 
\begin{equation}
    \log~EW = \alpha + \beta {\rm \log} L_{\rm 3000}, 
\end{equation}
where $EW$ denotes the equivalent width of the emission line and $L_{\rm 3000}$ represents the monochromatic luminosity at 3000 \AA. For each subgroup, we perform bootstrap resampling with 10,000 iterations (each iteration drawing a sample of quasars equal in size to the original subgroup) to derive the statistical uncertainties in $\alpha$ and $\beta$. Figures \ref{fig:BE_RQ} and \ref{fig:BE_RL} shows the fitting results for the RQ and RL quasars, respectively. The details of the relevant parameters are shown in Table \textcolor{blue}{\ref{table:fit}}. 

\begin{figure*}
	\centering
	\includegraphics[scale=0.6]{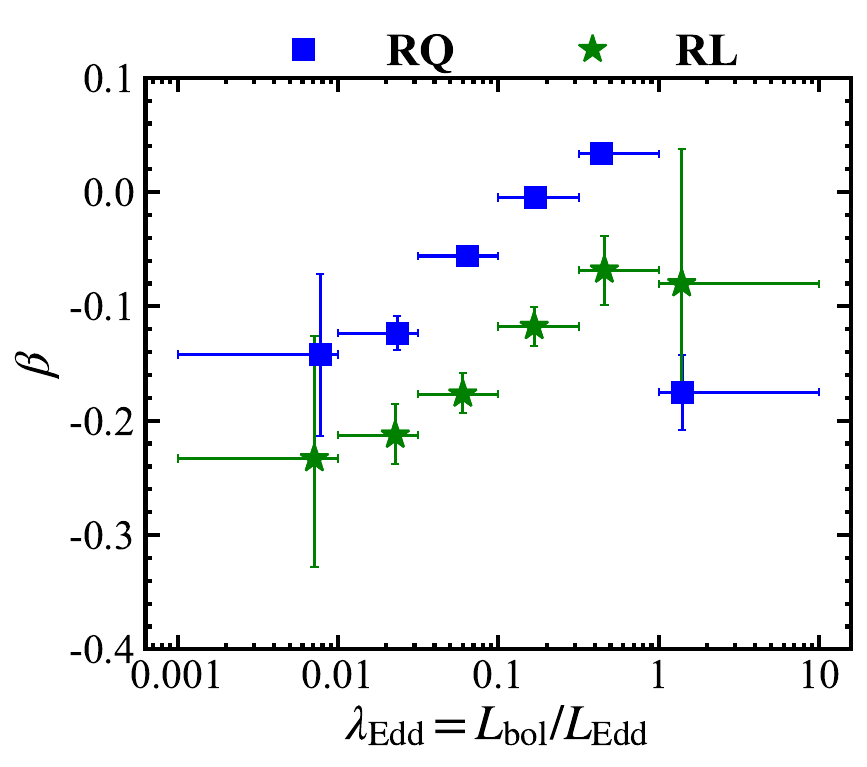}
	\caption{The dependence of Baldwin effect slope ($\beta$) on Eddington ratio ($\lambda_{\rm Edd}$). Blue squares denote RQ quasars, and green stars represent RL quasars. The horizontal error bars indicate the full width of each $\lambda_{\rm Edd}$ bin, and the data points are positioned at the median $\lambda_{\rm Edd}$ value within each bin.} At any fixed $\lambda_{\rm Edd}$, RL quasars consistently display steeper $\beta$ values than their RQ counterparts.
	\label{fig:beta}
\end{figure*}

As illustrated in Figure \textcolor{blue}{\ref{fig:beta}}, the Baldwin effect slope $\beta$ for both quasar samples increases monotonically with rising $\lambda_{\rm Edd}$, followed by a pronounced decline at the highest bin  ($\lambda_{\rm Edd}> 1$). Notably, at any fixed $\lambda_{\rm Edd}$, RL quasars consistently display steeper $\beta$ values than their RQ counterparts.

\begin{table*}
	\caption{Linear regression parameters for the Baldwin effect of \mgii\ in radio-quiet and radio-loud quasars.}
	\label{table:fit}
	\centering
	\begin{tabular}{ccccccc}
		\toprule[0.8pt]  
		\addlinespace[0.5pt] 
		\toprule[0.8pt]
		& \multicolumn{3}{c}{Radio-loud} & \multicolumn{3}{c}{Radio-quiet} \\
		\cmidrule(lr){2-4} \cmidrule(lr){5-7}
		${\rm log}~\lambda_{\rm Edd}$ & $N_{\rm QSO}$ & $\beta$ & $\alpha$ & $N_{\rm QSO}$ & $\beta$ & $\alpha$  \\
		\midrule
        (-3.0, -2.0] & 56 & $-0.233_{-0.095}^{+0.107}$ & $12.182_{-4.749}^{+4.246}$ & 156 & $-0.142_{-0.072}^{+0.070}$ & $8.051_{-3.082}^{+3.241}$ \\
		(-2.0, -1.5] & 576 & $-0.213_{-0.025}^{+0.027}$ & $11.214_{-1.231}^{+1.140}$ & 2546 & $-0.123_{-0.014}^{+0.015}$ & $7.202_{-0.647}^{+0.645}$ \\
		(-1.5, -1.0] & 1601 & $-0.177_{-0.017}^{+0.018}$ & $9.565_{-0.817}^{+0.761}$ & 11426 & $-0.056_{-0.007}^{+0.007}$ & $4.117_{-0.297}^{+0.304}$ \\
		(-1.0, -0.5] & 1690 & $-0.118_{-0.017}^{+0.017}$ & $6.842_{-0.767}^{+0.770}$ & 16539 & $-0.005_{-0.004}^{+0.005}$ & $1.709_{-0.207}^{+0.192}$ \\		
		(-0.5, 0] & 585 & $-0.068_{-0.031}^{+0.030}$ & $4.502_{-1.357}^{+1.405}$ & 5408 & $0.034_{-0.007}^{+0.007}$ & $-0.118_{-0.333}^{+0.316}$ \\
		(0, 1.0] & 73 & $-0.080_{-0.103}^{+0.118}$ & $5.088_{-5.413}^{+4.729}$ & 503 & $-0.175_{-0.033}^{+0.032}$ & $9.448_{-1.470}^{+1.514}$ \\	  
		\bottomrule
	\end{tabular}
\end{table*}

\section{DISCUSSIONS} \label{sec:discussion}

The observed positive correlation between $\beta$ and $\lambda_{\rm Edd}$, coupled with the systematically steeper $\beta$ exhibited by RL quasars at any given $\lambda_{\rm Edd}$, necessitates a coherent physical interpretation. Over the past several decades, multiple theoretical frameworks have been advanced to explain the Baldwin effect, however, a broad consensus has yet to be reached.

\begin{figure*}
	\centering
    \includegraphics[scale=0.55]{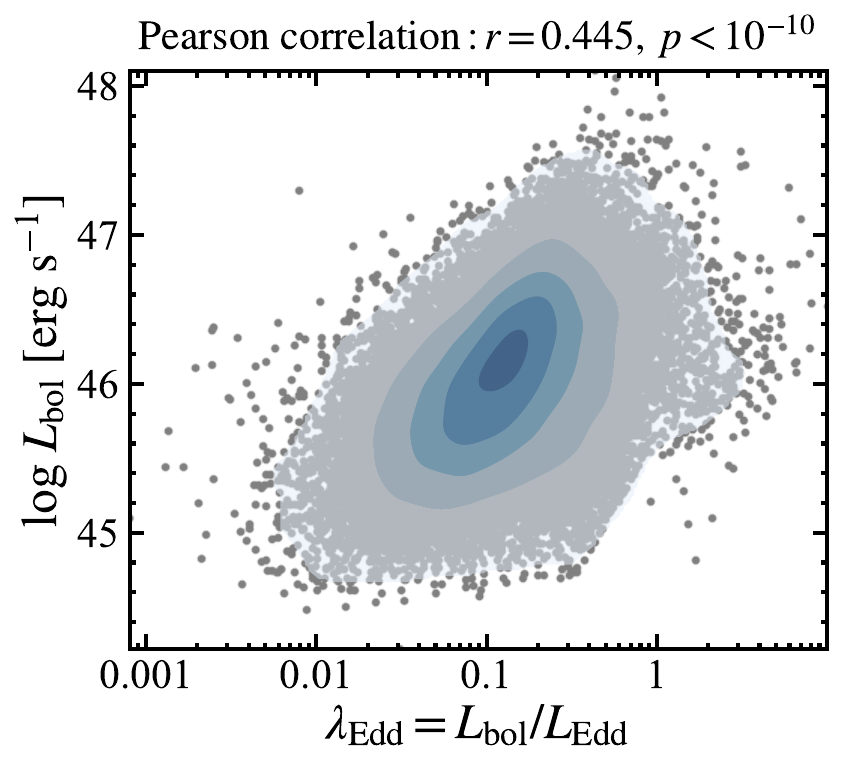}
	\caption{The bolometric luminosity versus Eddington ratio for our quasar sample. The Pearson correlation analysis yields a coefficient of $r = 0.445$ ($p < 10^{-10}$).}
	\label{fig:Lbol and Edd}
\end{figure*}

One class of models ascribes the Baldwin effect to luminosity-dependent variations in the shape of the ionizing continuum. Since AGN luminosity generally increases with $\lambda_{\rm Edd}$ (e.g., Figure \textcolor{blue}{\ref{fig:Lbol and Edd}}), observational and photoionization modeling studies consistently indicate that low-luminosity AGNs tend to possess harder extreme-UV continua (e.g., \citealt{Binette_1989, Wang_1998, Scott_2004}). A luminosity-driven softening of the ionizing continuum could suppress the EW of broad emission lines at higher luminosities, thereby reproducing the empirically established inverse EW–luminosity relationship (e.g., \citealt{Zheng_1993}). Nevertheless, our own photoionization simulations (details are given in the Appendix; Figure \textcolor{blue}{\ref{fig:cloudy1}a}) demonstrate that continuum spectral changes alone—absent concurrent modifications to other BLR properties—are insufficient to produce the observed significant variations in $\beta$.

An alternative mechanism invokes an additional line-emitting component associated with relativistic jets, potentially accounting for the steeper Baldwin effect slopes observed in RL quasars (e.g., \citealt{Victor_2016}). While this scenario may partially reconcile differences between RQ and RL populations, it fails to explain the positive $\beta-\lambda_{\rm Edd}$ correlation observed in our samples since high-$\lambda_{\rm Edd}$ quasars (regardless of radio loudness) consistently exhibit stronger disk-wind signatures (e.g., \citealt{Peng_2025}). Moreover, it offers no natural explanation for why RQ quasars—typically lacking prominent relativistic jets—nonetheless display a qualitatively similar $\beta-\lambda_{\rm Edd}$ evolutionary trend. Similarly, inverse Compton scattering models (e.g., \citealt{Xiao_2022}), which rely on jet-related processes to modify line emission, encounter analogous limitations.

A third class of models emphasizes the role of the BLR covering factor (e.g., \citealt{DongXB_2009}). In this framework, the EW of broad emission lines scales inversely with $\lambda_{\rm Edd}$ due to a decreasing covering factor at higher accretion rates—a prediction broadly consistent with the observed EW–$\lambda_{\rm Edd}$ anti-correlation. However, such models do not adequately reproduce the systematic dependence of the Baldwin effect slope on $\lambda_{\rm Edd}$ (Figure \textcolor{blue}{\ref{fig:cloudy1}b}). Crucially, they also fail to account for the key observational finding that, at a fixed $\lambda_{\rm Edd}$, RL quasars consistently exhibit steeper Baldwin effect slopes than their RQ counterparts. 

In light of these limitations, differences in host galaxy properties—such as stellar mass and cold gas content—may provide a more physically grounded explanation for the observed variations in the relationship between $\beta$ and $\lambda_{\rm Edd}$.  Historically, RL quasars have been reported to preferentially reside in massive, early-type elliptical galaxies—systems commonly described as ``red and dead'', exhibiting negligible ongoing star formation and low cold molecular gas reservoirs (e.g., \citealt{Best_2005, Sikora_2007, Heckman_2014}). In contrast, RQ quasars are more frequently associated with late-type disk galaxies or interacting/merging systems—environments rich in cold gas and actively forming stars (e.g., \citealt{Heckman_2014}). However, recent analyses based on the LoTSS survey challenge this interpretation. As demonstrated by \citealt{Igo_2024}, when selection biases—particularly those arising from sample incompleteness in stellar mass and radio luminosity—are rigorously accounted for, the incidence of radio AGN is found to be comparable across quiescent and star-forming galaxies.  Consequently, host galaxy properties appear to play a comparatively minor role in shaping the $\beta-\lambda_{\rm Edd}$ correlation.

It is now widely accepted that clouds in the BLR originate from condensations within radiation-driven winds (e.g., \citealt{1997ApJ...474...91M,2008ARA&A..46..475H}). Within this theoretical framework, sources with higher $\lambda_{\rm Edd}$ are predicted to launch more powerful outflows (e.g., \citealt{Peng_2025}), which subsequently entrain and compress larger volumes of gas—thereby yielding higher BLR gas densities. Conversely, quasars with lower $\lambda_{\rm Edd}$ or those residing in gas-poor environments are expected to launch weaker radiation-pressure–driven outflows. Moreover, during the evolutionary transition from high- to low-$\lambda_{\rm Edd}$ states, AGN feedback efficiently removes circumnuclear gas, leading to a measurable decrease in the BLR gas density. We propose that the gas density in the BLR is the principal physical determinant of the ``global'' Baldwin effect—the Baldwin effect slope $\beta$ exhibits a positive correlation with BLR gas density. This hypothesis is fully consistent with empirical trends: higher $\lambda_{\rm Edd}$ correlates with higher BLR gas density and yields a shallower $\beta$, whereas at fixed $\lambda_{\rm Edd}$, RL quasars display steeper $\beta$ values than RQ quasars when RL quasars typically reside in gas-poor hosts (e.g., \citealt{Best_2005, Sikora_2007, Heckman_2014}).

To assess whether BLR gas density alone can account for these observational patterns, we conducted a comprehensive suite of photoionization simulations by systematically controlling three key parameters, including the spectral shape of the ionizing continuum, the BLR covering factor, and the gas density. As illustrated in Figure \textcolor{blue}{\ref{fig:cloudy1}}, neither the continuum shape (Figure \textcolor{blue}{\ref{fig:cloudy1}a}) nor the covering factor (Figure \textcolor{blue}{\ref{fig:cloudy1}b}) produces a statistically significant change in the Baldwin effect slope $\beta$. In stark contrast, gas density (Figure \textcolor{blue}{\ref{fig:cloudy1}c}) exerts a dominant and monotonic influence: at a fixed ionizing spectrum, clouds with lower gas densities become increasingly over-ionized at high continuum luminosities, resulting in diminished line equivalent widths and, consequently, steeper $\beta$ values. These simulation results provide direct theoretical support for BLR gas density as the primary driver of the ``global'' Baldwin effect. 

It is generally understood that the ionizing continuum shape, covering factor, and gas density co-evolve as quasars age. Specifically, younger, higher-$\lambda_{\rm Edd}$ quasars tend to possess softer ionizing continua (e.g., \citealt{2012MNRAS.425..907J}), smaller covering factors, and higher BLR gas densities. To model this physically motivated coupled evolution, we performed an additional set of self-consistent photoionization simulations, wherein all three parameters were varied simultaneously across observationally constrained ranges. As shown in Figure \textcolor{blue}{\ref{fig:cloudy2}}, the resulting Baldwin effect slope $\beta$ displays a robust positive correlation with $\lambda_{\rm Edd}$, in excellent quantitative agreement with the observed trend presented in Figure \textcolor{blue}{\ref{fig:beta}}. Both Figures \textcolor{blue}{\ref{fig:cloudy1}} and \textcolor{blue}{\ref{fig:cloudy2}} consistently demonstrates that the BLR gas density is the primary determinant of the $\beta$–$\lambda_{\rm Edd}$ relation, while changes in the ionizing continuum shape and covering factor serve only as secondary modulators.

Taken together, our findings establish a coherent physical framework: the ``global'' Baldwin effect observed in the \mgii\ emission line is primarily governed by the gas density in the BLR, which is itself determined by the interplay between accretion physics (quantified by the $\lambda_{\rm Edd}$) and host galaxy properties, particularly the available cold gas reservoir. Consistent with this picture, quasars exhibiting higher $\lambda_{\rm Edd}$ possess denser BLR clouds and consequently display shallower Baldwin effect slopes $\beta$, a trend robustly observed across both RQ and RL quasars. At a given $\lambda_{\rm Edd}$, RL quasars systematically exhibit steeper $\beta$ values—a direct consequence of their typical residence in gas-depleted, massive early-type hosts, where diminished gas supply leads to lower BLR gas densities. This unified interpretation reconciles previously disparate observational trends—namely, the dependence of $\beta$ on both $\lambda_{\rm Edd}$ and radio loudness—and resolves a long-standing ambiguity regarding the physical origin of the Baldwin effect. Crucially, it elevates the Baldwin effect from a phenomenological correlation to a quantitative diagnostic tool for probing AGN accretion states and coevolutionary processes between supermassive black holes and their host galaxies.

\section{summary}

This study investigates the physical origin of the \mgii\ broad emission line Baldwin effect using a large sample of 41,159 radio quasars, based on the SDSS DR16Q-LoTSS DR2 matched sample from the radio–optical identification catalog of \citet{Hardcastle_2023}. This sample includes 36,578 RQ quasars with radio-loudness $R\leq1.47$ and 4,581 RL quasars wiht $R>1.47$.

Key findings reveal that the Baldwin effect slope $\beta$ (defined by $\log \text{\rm EW} = \alpha + \beta \log L_{\rm 3000}$, where $L_{\rm3000}$ is the 3000 \AA\ continuum luminosity) exhibits a positive correlation with the Eddington ratio $\lambda_{\text{Edd}}$ in both RQ and RL populations. Additionally, at fixed $\lambda_{\text{Edd}}$, RL quasars consistently show steeper $\beta$ values than RQ quasars.

Photoionization simulations demonstrate that$\beta$ is primarily governed by the gas density in the broad-line region: lower gas densities result in steeper $\beta$. This density-driven mechanism links the Baldwin effect to AGN evolutionary processes: higher $\lambda_{\text{Edd}}$ drives stronger disk winds, compressing gas to form denser BLRs and shallower $\beta$. 

These results suggest that the gas density of broad-line region plays a central role in driving the ``global" Baldwin effect, providing a new perspective on AGN accretion states and the coevolution of supermassive black holes with their host galaxies.

\begin{acknowledgments}
This work is supported by the Guangxi Natural Science Foundation (2024GXNSFDA010069; 2026GXNSFGB00640005), the National Natural Science Foundation of China (12473011). 
Funding for the Sloan Digital Sky Survey V has been provided by the Alfred P. Sloan Foundation, the Heising-Simons Foundation, the National Science Foundation, and the Participating Institutions. SDSS acknowledges support and resources from the Center for High-Performance Computing at the University of Utah. SDSS telescopes are located at Apache Point Observatory, funded by the Astrophysical Research Consortium and operated by New Mexico State University, and at Las Campanas Observatory, operated by the Carnegie Institution for Science. The SDSS web site is \url{www.sdss.org}.
SDSS is managed by the Astrophysical Research Consortium for the Participating Institutions of the SDSS Collaboration, including Caltech, The Carnegie Institution for Science, Chilean National Time Allocation Committee (CNTAC) ratified researchers, The Flatiron Institute, the Gotham Participation Group, Harvard University, Heidelberg University, The Johns Hopkins University, L'Ecole polytechnique f\'{e}d\'{e}rale de Lausanne (EPFL), Leibniz-Institut f\"{u}r Astrophysik Potsdam (AIP), Max-Planck-Institut f\"{u}r Astronomie (MPIA Heidelberg), Max-Planck-Institut f\"{u}r Extraterrestrische Physik (MPE), Nanjing University, National Astronomical Observatories of China (NAOC), New Mexico State University, The Ohio State University, Pennsylvania State University, Smithsonian Astrophysical Observatory, Space Telescope Science Institute (STScI), the Stellar Astrophysics Participation Group, Universidad Nacional Aut\'{o}noma de M\'{e}xico, University of Arizona, University of Colorado Boulder, University of Illinois at Urbana-Champaign, University of Toronto, University of Utah, University of Virginia, Yale University, and Yunnan University.
\end{acknowledgments}

\facilities{SDSS, LOFAR}

\software{astropy \citep{2013A&A...558A..33A,2018AJ....156..123A,2022ApJ...935..167A}
}

\begin{appendix}
\renewcommand{\thesection}{Appendix}

\section{Photoionization simulation}  \label{subsec:simulate}
To investigate the underlying physical mechanisms governing the $\beta$–$\lambda_{\rm Edd}$ in both RQ and RL quasar samples, we employed the photoionization code \textsc{Cloudy} package (version C25.00; \citealt{2025RMxAA..61..120G}) to construct self-consistent photoionization models of the BLR in AGN. The configuration of these models is described in detail below.

(1) For the incident spectral energy distribution (SED), we adopted the Eddington-ratio-dependent AGN continuum templates from \cite{2012MNRAS.425..907J}, which are empirically calibrated and span three accretion regimes: low, intermediate, and high Eddington ratios (hereafter denoted as Low-Edd, Mid-Edd, and High-Edd, respectively). These templates exhibit systematic differences—most notably in the amplitude of the ultraviolet (UV) bump and the strength of the soft X-ray excess—reflecting physically motivated variations in the structure and emission properties of the accretion disk and corona as a function of accretion rate.

(2) Each \textsc{Cloudy} simulation models a single spherical BLR cloud illuminated by the AGN continuum. We adopt spherical geometry (\texttt{sphere}) and specify the incident radiation field using the \texttt{intensity total} command, which defines the bolometric flux incident on the cloud's illuminated surface. To anchor the simulations to the observationally accessible quantity $L_{3000}$, we compute the radial distance $r$ between the central ionizing source and the cloud—thereby converting the source's total (bolometric) luminosity into the local radiation intensity required by \textsc{Cloudy}. This conversion relies on the empirically calibrated radius–luminosity (R–L) relation \citep{10.1093/mnras/stad1224}:
\begin{equation}
    \log r = \alpha\,(\log\,L_{3000} - 45) + 17.48,
    \label{eq:rl}
\end{equation}
where $r$ is expressed in cm, $L_{3000}$ denotes the monochromatic luminosity at 3000 \AA\ in $\mathrm{erg\,s^{-1}}$, and the R–L slope $\alpha$ is treated as a free parameter within $0.39 \pm 0.08$. For each input $L_{3000}$, we first estimate the bolometric luminosity as $L_{\mathrm{bol}} \approx 5\,L_{3000}$, then determine the characteristic BLR radius $r$ via Equation (\ref{eq:rl}), and finally compute the incident intensity as $I = L_{\mathrm{bol}} / (4\pi r^2)$, which is supplied to \textsc{Cloudy} through the \texttt{intensity total} command.

We explore hydrogen number densities in the range $\log\,n_{\mathrm{H}} = (9.5$–$11.5$), bracketing values inferred from the Locally Optimally Emitting Clouds (LOC) model \citep{1995ApJ...455L.119B}. The hydrogen column density is fixed at $\log\,N_{\mathrm{H}} = 22$–$23$, ensuring optical thickness to ionizing photons and consistency with the partially ionized zone—where \mgii\ emission is predominantly generated. The covering factor (CF) is varied from 0.1 to 0.3, and the gas-phase metallicity is set to $Z = 5\,Z_{\odot}$, motivated by spectroscopic studies of BLRs indicating systematically enhanced metallicities in AGN \citep{1999ARA&A..37..487H}. A microturbulent velocity of $v_{\mathrm{turb}} = 500\ \mathrm{km\,s^{-1}}$ is adopted—typical of BLR kinematics—and a filling factor of $\mathrm{FF} = 0.2$ is assumed. 

\begin{figure*}
    \centering
    \begin{minipage}{0.32\linewidth}\centering\textbf{(a)}\\
        \includegraphics[width=\linewidth]{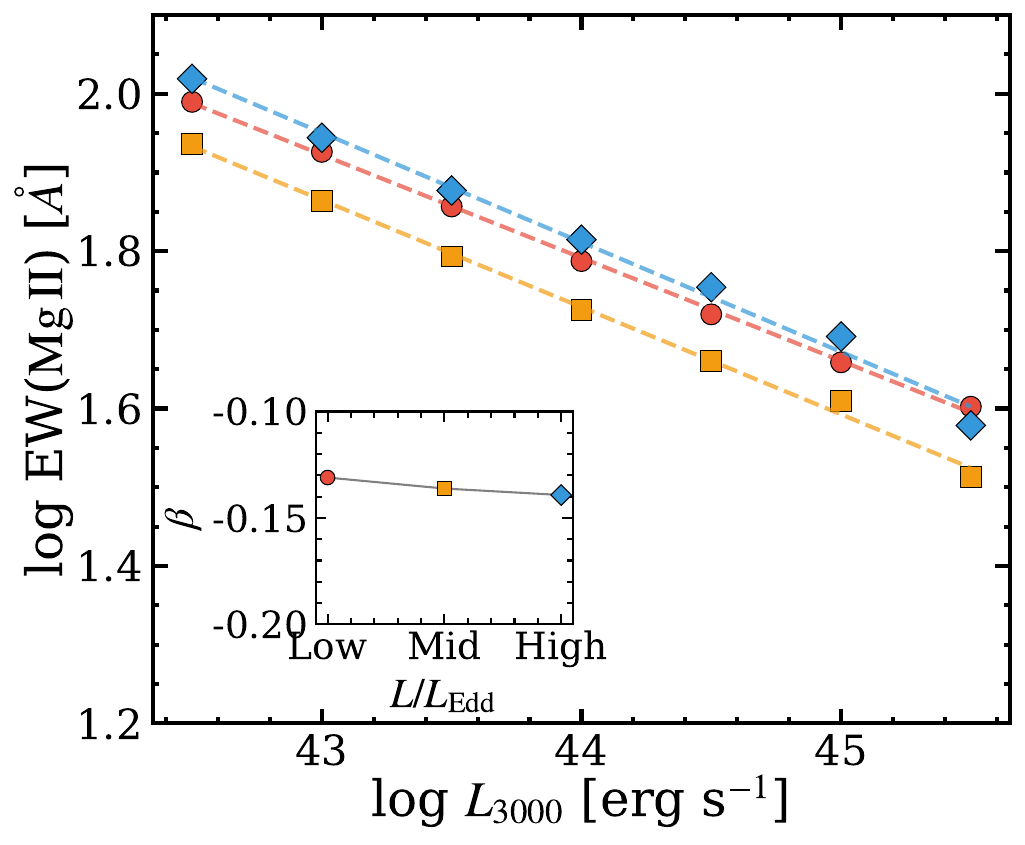}
    \end{minipage}
    \begin{minipage}{0.32\linewidth}\centering\textbf{(b)}\\
        \includegraphics[width=\linewidth]{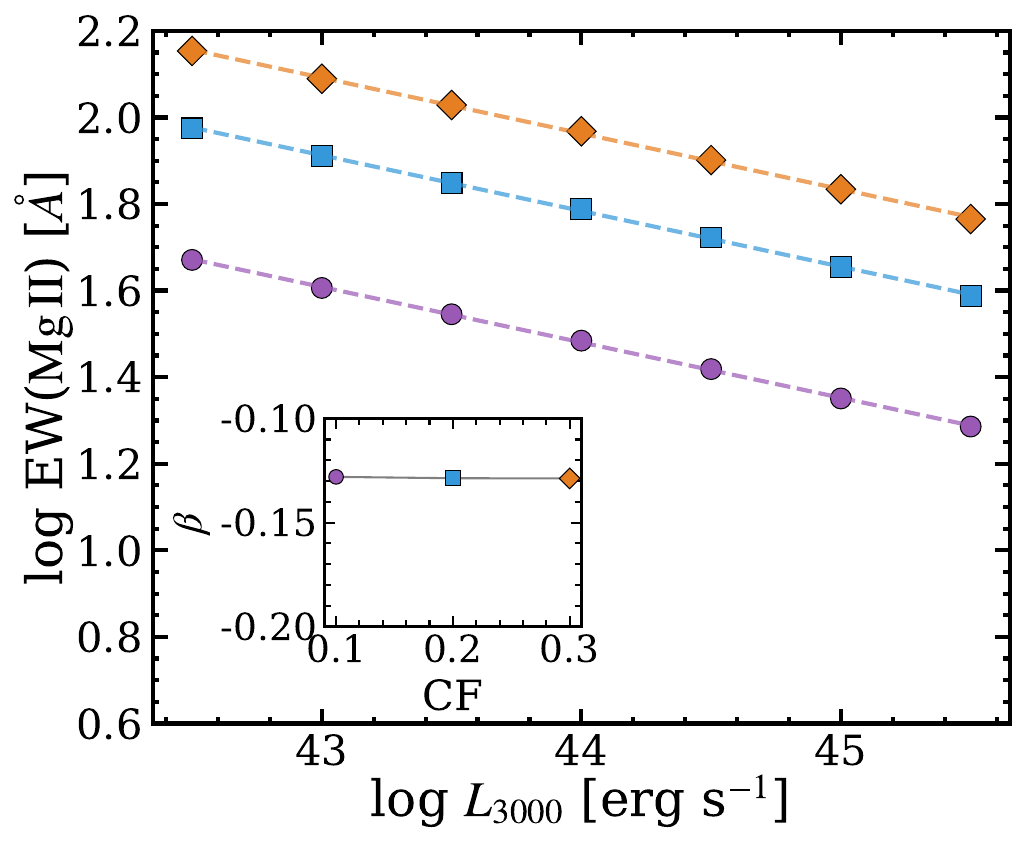}
    \end{minipage}
    \begin{minipage}{0.32\linewidth}\centering\textbf{(c)}\\
        \includegraphics[width=\linewidth]{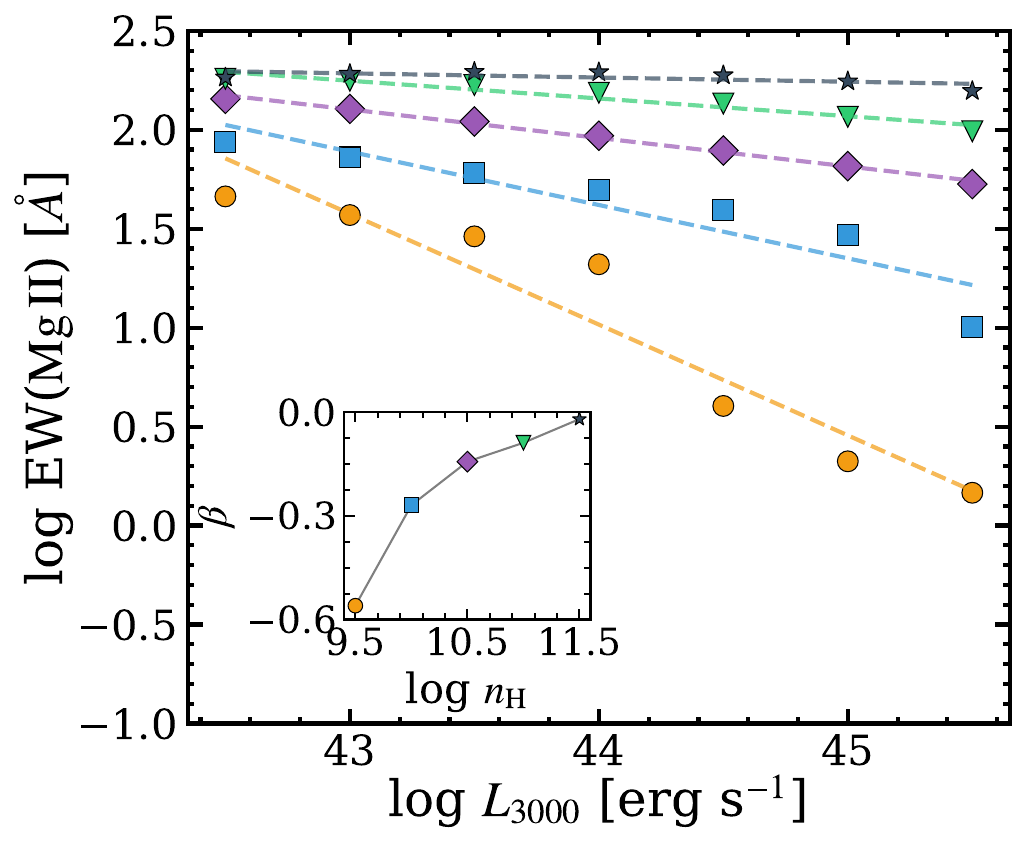}
    \end{minipage}      
        \caption{Cloudy photoionization simulations were performed to examine the Baldwin effect of \mgii\ emission lines in three controlled BLR parameter configurations: (a) variations of the ionizing continuum alone, with the BLR covering factor and gas density held fixed; (b) variations of the BLR covering factor only, with the ionizing continuum and gas density held constant; and (c) variations of the gas density only, with the ionizing continuum and BLR covering factor held constant. In each panel, the inset displays the dependence of the Baldwin effect slope $\beta$ on the respective varied parameter. Obviously, the results indicate that $\beta$ is most sensitive to BLR gas density: higher densities yield shallower (i.e., less negative) $\beta$ values, whereas variations in the ionizing continuum or BLR covering factor produce negligible changes in $\beta$.}
    \label{fig:cloudy1}
\end{figure*}

Figure \textcolor{blue}{\ref{fig:cloudy1}} presents the \textsc{Cloudy} simulation results investigating the Baldwin effect of \mgii\ emission lines, systematically varying key physical parameters across three distinct model configurations:  
\begin{itemize}
    \item \textbf{Configuration A}: The incident SED is varied among the three Eddington-ratio-dependent templates from \cite{2012MNRAS.425..907J} — Low-Edd, Mid-Edd, and High-Edd — while the covering factor (CF) and hydrogen number density are held fixed at $\mathrm{CF} = 0.2$ and $\log\,n_{\mathrm{H}} = 10.0$, respectively. The results are presented in Figure \textcolor{blue}{\ref{fig:cloudy1}a}, showing that the Baldwin effect slope $\beta$ remains essentially unchanged when varying the incident SED.
    \item \textbf{Configuration B}: The covering factor is varied over the range $0.1$–$0.3$, adopting the classical AGN SED (Mid-Edd template from \citealt{2012MNRAS.425..907J}) and fixing $\log\,n_{\mathrm{H}} = 10.0$. As shown in Figure \textcolor{blue}{\ref{fig:cloudy1}b}, the Baldwin effect slope $\beta$ exhibits no significant dependence on the covering factor.
    \item \textbf{Configuration C}: The hydrogen number density is varied over $\log\,n_{\mathrm{H}} = 9.5$–$11.5$, adopting the Mid-Edd template from Jin et al. (2012) and fixing $\mathrm{CF} = 0.2$. Figure \textcolor{blue}{\ref{fig:cloudy1}c} demonstrates that the Baldwin effect slope $\beta$ is sensitive to variations in the hydrogen number density.
\end{itemize}  

To further investigate how varying Eddington ratios influence the slope of the Baldwin effect, we constructed three physically motivated configurations that simulate the co-evolution of BLR properties as quasars age. These configurations are based on the templates of \cite{2012MNRAS.425..907J} and represent three distinct evolutionary stages:

(i) an old, low-$\lambda_{\rm Edd}$ stage (Low-Edd template, CF = 0.3, $\log n_{\rm H} = 9.5$), where the ionizing continuum is harder, the covering factor is larger, and the BLR gas density is lower due to reduced outflow activity;

(ii) a mature, moderate-$\lambda_{\rm Edd}$ stage (Mid-Edd template, CF = 0.2, $\log n_{\rm H} = 10.5$); and

(iii) a young, high-$\lambda_{\rm Edd}$ stage, characterized by a soft ionizing continuum (High-Edd template), a small covering factor (CF = 0.1), and a high BLR gas density ($\log n_{\rm H} = 11.5$), reflecting strong disk winds that compress the BLR gas.

Our results indicate that the derived Baldwin effect slope increases with increasing Eddington ratio $\lambda_{\rm Edd}$, as illustrated in Figure 7.

\begin{figure*}
    \centering
    \includegraphics[width=0.5\linewidth]{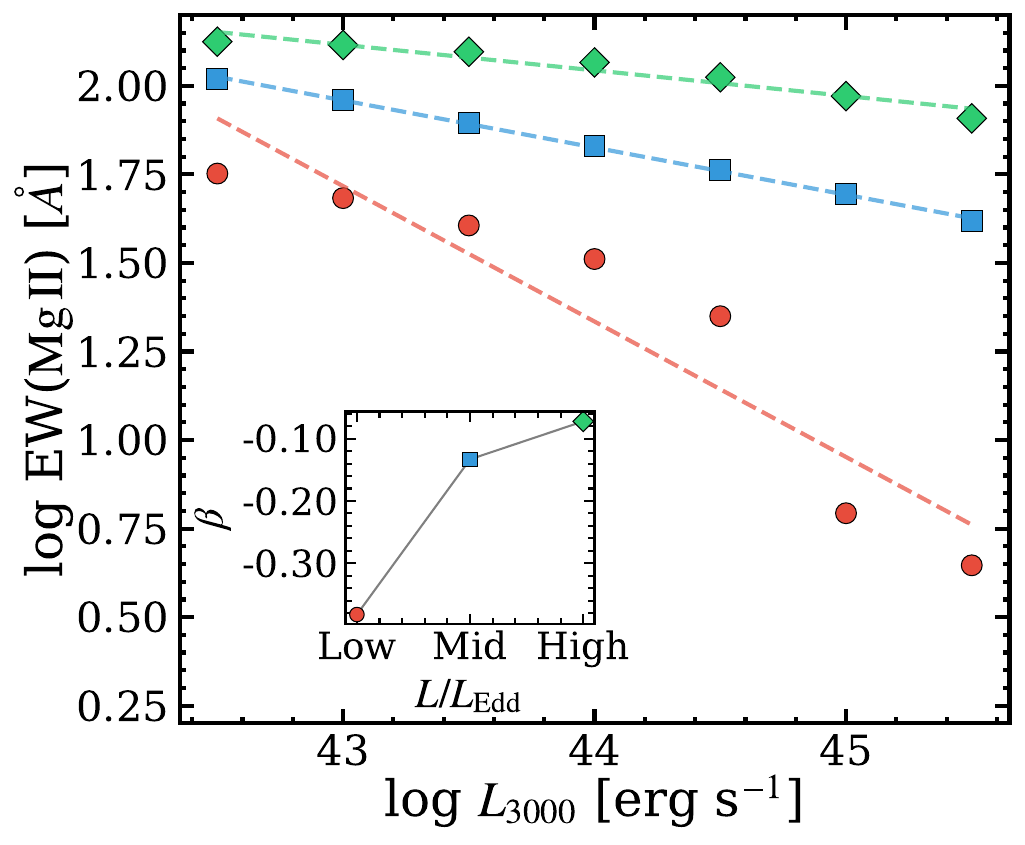}
    \caption{Cloudy simulations were conducted to investigate the Baldwin effect of \mgii\ emission lines in three physically motivated configurations: (1) a low Eddington ratio regime, associated with a hard ionizing SED, CF = 0.3, and $\log n_{\rm H} = 9.5$ (red circles); (2) a moderate Eddington ratio regime, featuring an intermediate ionizing SED, CF = 0.2, and $\log n_{\rm H} = 10.5$ (blue squares); and (3) a high Eddington ratio regime, characterized by a soft ionizing spectral energy distribution (SED), a BLR covering factor of CF = 0.1, and a hydrogen number density of $\log n_{\rm H} = 11.5$ (green rhombus). The inset panel displays the variation of the Baldwin effect slope $\beta$ across three $\lambda_{\rm Edd}$ configurations.}
    \label{fig:cloudy2}
\end{figure*}

\end{appendix}

\bibliography{refs}{}
\bibliographystyle{aasjournal}



\end{document}